\documentclass[aps,12pts,prb,amsmath,amssymb,preprint,showkeys,showpacs]{revtex4-1}
\usepackage{graphicx}
\usepackage{bm}
\usepackage{mathptmx}
\usepackage{epstopdf}
\usepackage{booktabs}
\usepackage{multirow}
\usepackage{hhline}
\usepackage{dcolumn}
\usepackage[T1]{fontenc} 
\usepackage{amssymb,amsmath,graphics,graphicx,float,braket}
\usepackage{epstopdf}
\usepackage{array,multirow,setspace}
\usepackage{subfig}
\usepackage[autostyle]{csquotes}
\usepackage{dcolumn}
\usepackage{bm}
\usepackage{booktabs}
\usepackage{physics}\usepackage{subscript}
\usepackage{soul}
\usepackage[utf8x]{inputenc} 
\usepackage{verbatim}
\usepackage{hyperref}
\usepackage{url}

\begin{document}
\title{A general rule for predicting the magnetic moment of Cobalt-based Heusler compounds using compressed sensing and density functional theory}
\author{Satadeep Bhattacharjee$^{1}$}
\email{s.bhattacharjee@ikst.res.in}
\affiliation{Indo-Korea Science and Technology Center (IKST), Bangalore, India}
\author{Seung-Cheol Lee$^2$}
\email{leesc@kist.re.kr}
\affiliation{Electronic Materials Research Center, Korea Institute of Science $\&$ Technology, Korea}
\begin{abstract} We propose a general rule for estimating the magnetic moments of Co$ 2$(cobalt)-based Heusler alloys, especially when doped with late transition metals. We come up with a descriptor that can characterise both pure Co$_2$YZ compounds and the doped ones with the chemical formula Co$_2$Y$_{1-x}$M$_x$Z (M is the dopant) using online data for magnetic moments of Heusler alloys with Co$_2$YZ structure and compressive sensing approach. The newly proposed descriptor not only depends on the number of valence electrons of the compound also it depends on the number of unoccupied d-electrons in the doping site. A comparison of the performance of the proposed descriptor and the Slater-Pauling rule is made.  Unlike the Slater-Pauling rule, which is only effective for half-metallic Heusler compounds, our machine-learning approach is more generic since it applies to any Co$_2$YZ Heusler compounds, regardless of whether they are half-metals or not. We use this new rule to estimate the magnetic moments of a few yet-to-be-discovered Heusler compounds and compare the results to density functional theory (DFT) based calculations. Finally, we use DFT and machine learning investigations to prove their stability.
\end{abstract}

\maketitle

\section{Introduction}
Heusler compounds are important because of their potential application in the field of spintronics~\cite{spin1,felser2007spintronics,review-recent,spin-l}. Depending on the nature of of magnetic sub-lattices an Heusler alloy is called either half Heusler compounds or full Heulser compounds~\cite{galanakis2002slater,full}. Half Heusler compounds have the chemical formula XYZ while the full Heusler alloys have the chemical formula X$_2$YZ. Here X and Y are the transition metal element while Z is an sp-element. Full Heusler compounds can have either L2$_1$ structure with atomic ordering of X[8c(1/4,1/4,1/4)], Y[4a(0,0,0)], Z[4b(1/2,1/2,1/2)], or inverse Heusler alloys (also known as XA structure) where X atoms occupy 4a (0,0,0) and 4c (1/4,1/4,1/4) positions while Y and Z atoms are found in 4b (1/2,1/2,1/2) and 4d (3/4,3/4,3/4) respectively~\cite{new}. Also, there could be of martensite phase with D022~\cite{D1,D2,D3} structure.

New stable Heusler alloys with relatively large magnetic moments and Curie temperatures are of great interest to the materials research community working on related materials. To predict the magnetic moment of a new Heusler compound one needs a guideline. Counting the number of valance electrons is a simple way to estimate the magnetic moments of several Heusler compounds using the much celebrated Slater-Pauling rule~\cite{SP1,SP2,SP3,Phivos}. For full Heusler alloys, this rule states that the magnetic moment, $M$, per formula unit can be related to the number of valence electrons, $N_v$ via $M=N_v-24$. For half Heusler compounds, the same rule predicts the magnetic per formula unit to be $M=N_v-18$. Despite of such a simplicity, Slater-Pauling (SP) rule  is a great success. 

It has been seen, nonetheless, that not all Heusler alloys agree with this straightforward counting criterion. The SP rule typically works for Heusler alloys with cubic structure and perfect stoichometry and half-metallic nature. Away from full stoichiometry and half-metallic nature, variations from the SP rule are frequently encountered. Particularly when the Heusler compounds have lower symmetry because of non-stoichiometry.

Even for the stoichometric one with cubic structures, the SP rule often fails for Heusler alloys with late transition metals, which is quite remarkable phenomenon as observed for example in inverse Heusler alloys such as Fe$_2$CuAl~\cite{Late1}. 
\par
To obtain new Heusler alloys with significant magnetic moment, a simple strategy is to alloy the existing ones with other elements. The alloying can be done on either of X, Y or Z sites.
In the present work, we restrict the doping site to be Y-site or more precisely we consider the ones with the chemical composition Co$ 2$Y$_{1-x}$M$_x$Z,(M is the dopant). 

In this work, we use machine learning techniques to critically examine the SP rule and propose a new general rule that can predict the magnetic moments Co$_2$ based full Heusler alloys~\cite{khandy}of the form of both Co$_2$YZ and Co$_2$Y$_{1-x}$M$_x$Z in an effective manner.  Using a small dataset of the magnetic moments of Co$_2$ based  Heusler alloys with L2$_1$ symmetry and using compressed sensing method, we obtain a suitable low-dimensional descriptor for the magnetic moment. We particularly point out that the incompatibility of the SP rule in the presence of late transition metals can be resolved by introducing certain corrections to the original SP rule in terms number of unoccupied d-states of the metal site that contains the late transition metal.
It is worth noting that the importance of unoccupied d-states was originally noticed in the fundamental work of K\"ubler \textit{et al.}~\cite{kub}, which correlates the magnetic moment to the localisation of the unoccupied minority d-band at the Y-site.
\par To show the performance of the current approach, we adopt the following strategy: We obtain a three dimensional descriptor for magnetic moment of Heusler alloys with chemical formula Co$_2$YZ and Co$_2$Y$_{1-x}$M$_x$Z with cubic structure. Next, from the three-dimensional descriptor we construct an effective two-dimensional descriptor which we refer as our general rule. To further test the validity of such descriptor, we calculate the magnetic moments of few doped compounds i,e Co$_2$Y$_{1-x}$M$_x$Z material with such descriptor and compare them with the values obtained from the  density functional theory (DFT) based method as well as from the standard SP rule. Finally, the  stability of such compounds are again evaluated using similar compressed sensing approach.  The stability of these compounds are then re-verified with DFT approach.
\par The manuscript is organized as follows: Following a brief description of the computational methods used in the study, we describe the data-set used, followed by the results and discussion section, in which we formulate our general rule and demonstrate its predictability in comparison to both the standard Slater-Pauling rule and the generalised one. Finally, we reach conclusions.
                      
\section{Computational methods}
\subsection{Compressed sensing}
We use the compressed sensing approach based method namely Sure Independence Screening and Sparsifying Operator (SISSO)\cite{ouyang2018sisso,xu2020data,ouyang2019simultaneous}, which works in iterative manner. One of the advantage of this method is that one can obtain mathematical formulation of the desired descriptors~\cite{nayak,acosta2018analysis,fat}. Within the current approach, the properties of interest, $P_1$,$P_2$....$P_N$, can be expressed as linear functions of candidate features, ${\bf d}_1$,${\bf d}_2$....${\bf d}_M$.

The model and the descriptors are obtained by minimizing;
\begin{equation}
\underset{C}{\operatorname{arg min}} ||\bf P-\bf D \bf c||^2_2+\lambda||C||_0
\label{CS}
\end{equation}
Where $||\bf c||$ is the $l_0$ norm of c. D is a $N\times M$ matrix known as sensing matrix.
The vector $\bf c$ has $\Omega$-non-zero components known as sparsity. We construct the feature space with the following operators $H^{(m)}\equiv \{(+)(-)(*)(/)(sqrt)(^2)(^3) (exp)(cbrt)\}$ in the case magnetic moment while for the case formation energy we use a large set operators given by $H^{(m)}\equiv\{(+)(-)(*)(/)(sqrt)(^2)(^3)(^-1)(log)(exp)(exp-)\}$.
Within SISSO approach, the candidate features are constructed as non-linear functions of the primary features $\Phi_0$.
The feature space $\Phi_n$ (n=1,2,3..) is made by starting from a set of primary features given by $\Phi_0$ and then combining the features generated in an iterative way using the operators mentioned in $H^{(m)}$.
\subsection{Density functional theory~(DFT)}
The electronic structure calculations were performed using first-principles methods within the frame-work of Density Functional Theory (DFT) with Perdew-Burke Ernzerhof exchange correlation energy functional~\cite{pbe} based on a generalized gradient approximation. We used a projector augmented wave method as implemented in Vienna \textit{ab-initio} simulation package (\textsl{VASP})~\cite{vasp}.  Kohn-Sham wave functions of the valence electrons were expanded in plane wave basis with energy cut-off of 450 eV. The augmentation charge cut-off was set to be 627.1 eV. The Brillouin zone sampling was carried out using Monkhorst Pack grid of $\mathrm{7\times 7\times 7}$ k-points. Ionic relaxation was performed using conjugate-gradient method, until forces were reduced to within 0.01 eV/Angstrom. 
\section{The Data-set}
\begin{figure*}[h!]
\centering
\includegraphics[width=0.85\columnwidth,height=150mm]{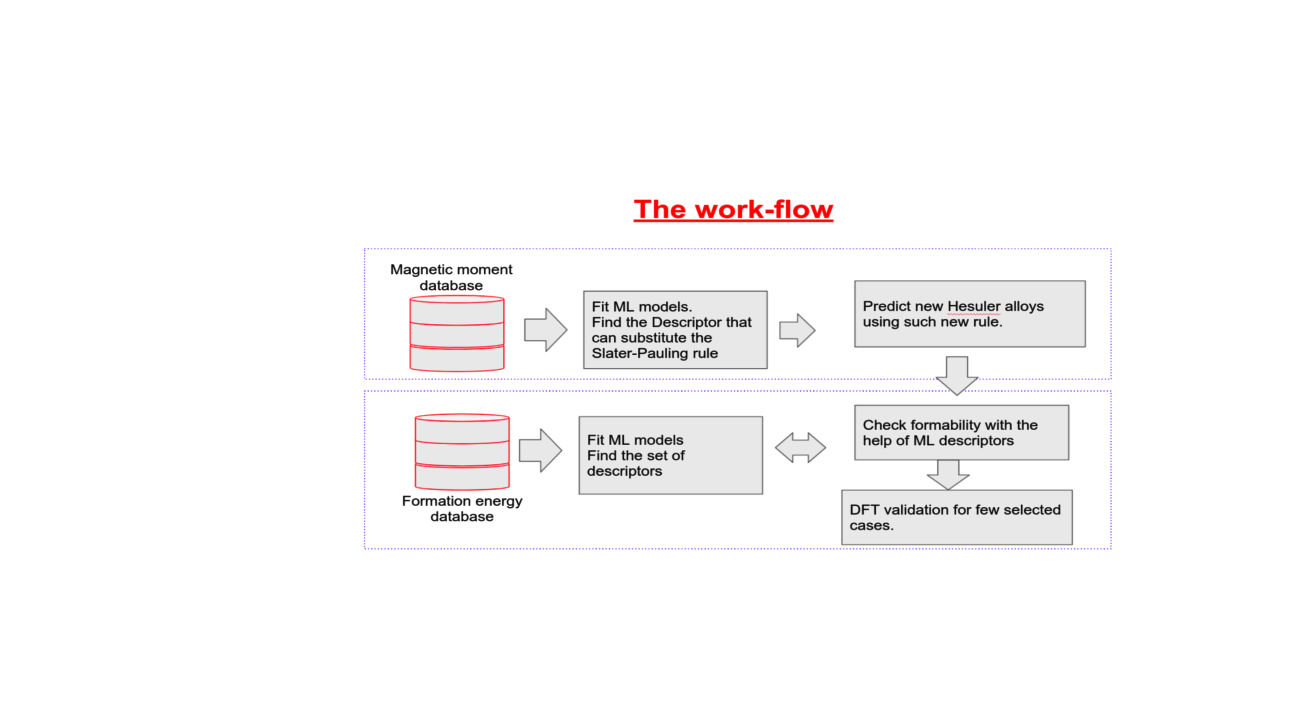}
\caption{The schematics of the workflow to predict new Heusler alloys using a new general rule proposed by us. }
\label{flow}
\end{figure*}

\begin{figure*}[h!]
\subfloat[Formation Energy]{
		\includegraphics[width=0.5\columnwidth, height=60mm]{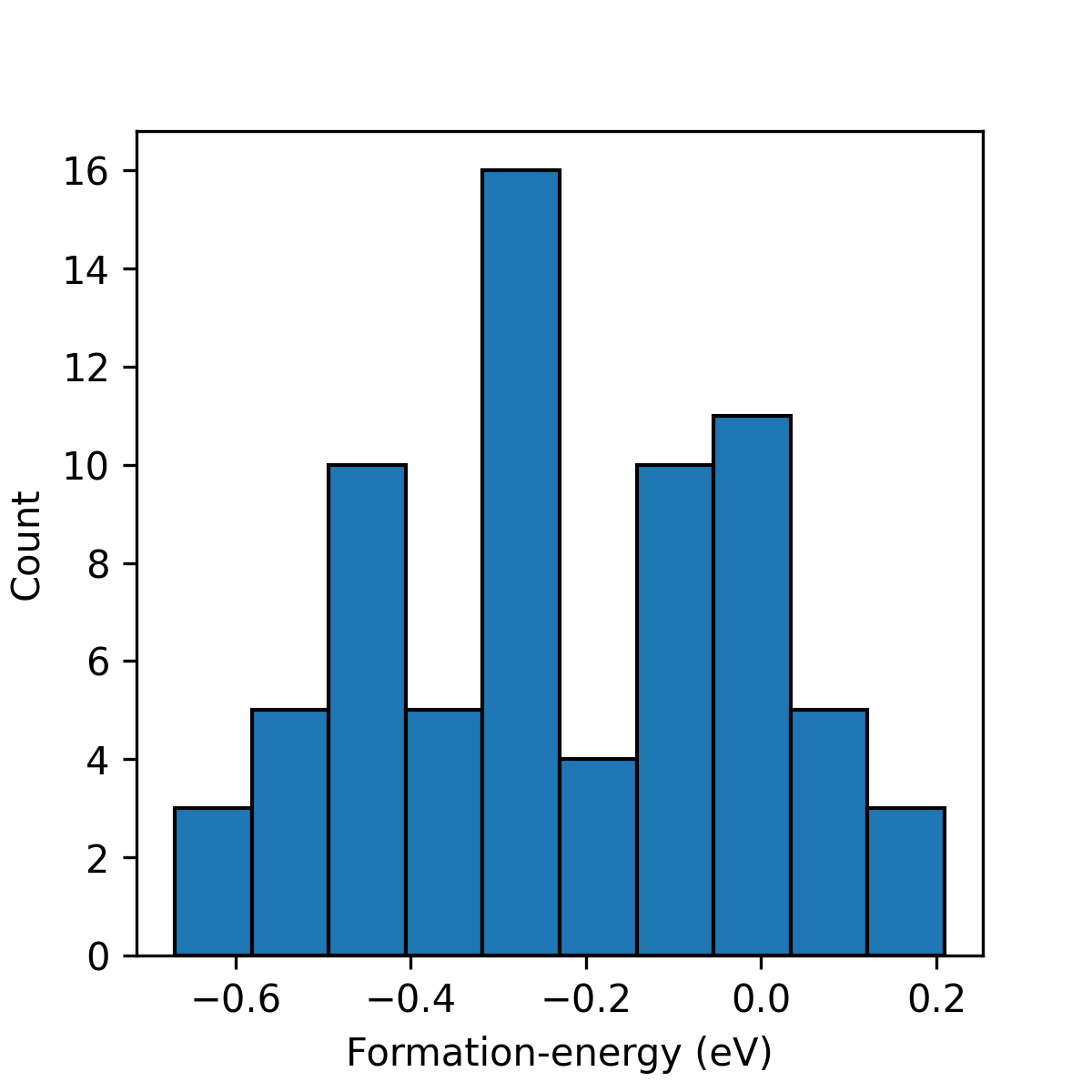}}
	\subfloat[Magnetic moment]{
		\includegraphics[width=0.5\columnwidth, height=60mm]{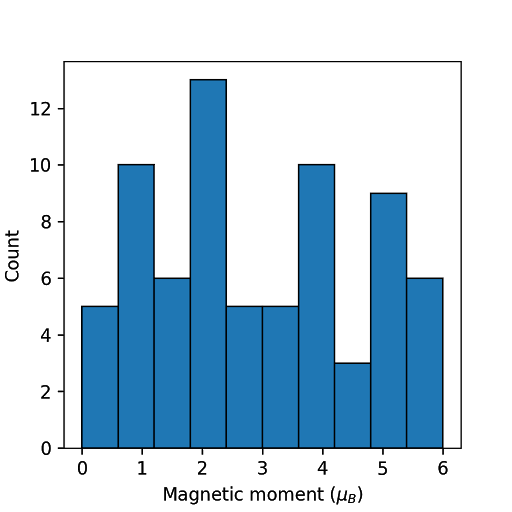}}
	\caption{Distribution of (a)formation energy (b)magnetic moment data.}
	\label{2D1}
\end{figure*}

The flow of this manuscript is  shown in the Fig.\ref{flow}, which shows how the general rule that we propose may be employed to identify novel Heusler compounds. 
We used data extracted from the Heusler alloy database C-spin~\cite{albama}. We dataset contains the, structure, magnetic moment and formation energies of Co$_2$YZ type heusler alloys. Our data set consists of 72 Co$_2$-based full Heusler alloys. The distribution of the magnetic moments and formation energies of these materials are shown in the Fig\ref{2D1}. The formation energy data has the mean at -0.22 eV and the standard deviation of 0.21, while the magnetic moment distribution has mean of 2.87 $\mu_B$  and standard deviation of 1.73. The data set contains both L2$_1$ (space group 225)  and $D022$ (space group 139) types.
To obtain a proper ML model for the Heusler compounds doped on Y-site, we also considered few new entries which are obtained by performing DFT calculations on the pure Heusler compounds doped with TM atoms. We consider only seven new entries, out of which five correspond to 25\%(Co$_2$Y$_{0.75}$M$_{0.25}$Z) doping while remaining two correspond to 50\% (Co$_2$Y$_{0.5}$M$_{0.5}$Z) doping. To obtain these data we consider the conventional cell for the L2$_1$ structure with 16 atoms and substitute either one (for 25\% doping) or two (for 50\% doping) Y atoms with the dopant atom. The relaxations are performed as described above.
\section{Results and discussions}
\subsection{Prediction of magnetic moments}
Within the realm of Slater-Pauling rule, the magnetic moment is related to the number of valence electrons. The magnetic moment is given by, $M_{SP}=n_\uparrow-n_\downarrow=N_v-2n_\downarrow$. Where $N_v=n_\uparrow+n_\downarrow$ is the number of the valence electrons. As shown by Galanakis \textit{et al.}~\cite{galanakis2002slater}, $n_\downarrow=12$ out of which 8 are d-electrons, 3 sp-electrons and 1 s-electron, thus giving the $M_{SP}=N_v-24$ rule for the Co$_2$YZ compounds.

 Using the magnetic moment dataset described above we first look for appropriate ML-model via the compressed sensing approach within the framework of SISSO method. For this purpose, we choose only the ones with L2$_1$ symmetry along side with the seven doped ones. This choice comes from the fact that we are interested in developing an empirical formula for the magnetic moment for the Heusler compounds that are \textit{cubic} unlike the case of D022 which are tetragonal. The newly added seven compounds which have mixed stoichiometry at Y site have slightly lower symmetry than L2$_1$, however they are still cubic (space group: $Pm\bar3m$).
 
 The first objective is to find a magnetic moment descriptor that can be used instead of the Slater-Pauling rule.  
 
\begin{figure}[hbt!]
    \centering
    \subfloat[$ $]{\includegraphics[scale=0.8]{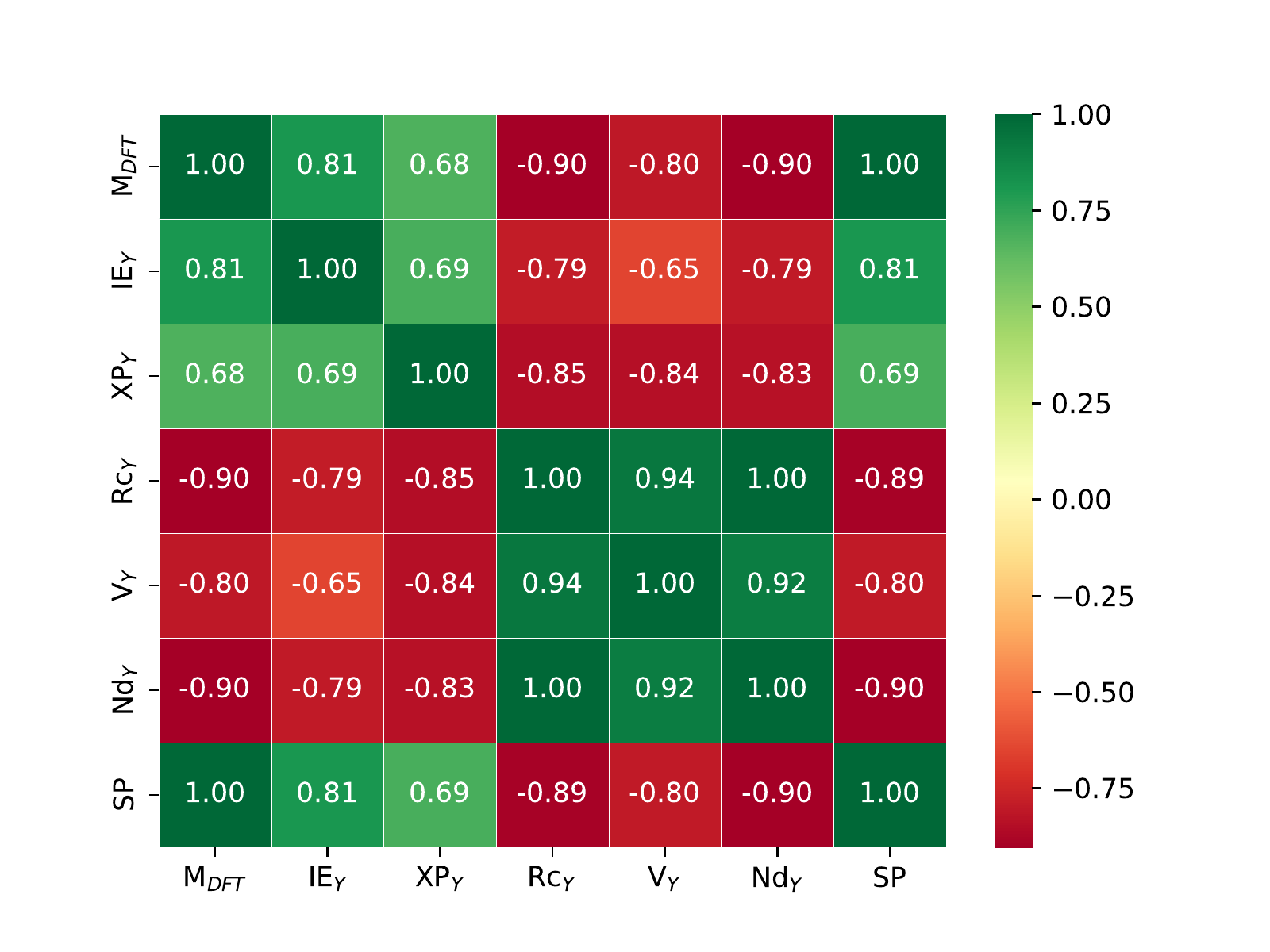}}\\
    \subfloat[$ $]{\includegraphics[scale=0.8]{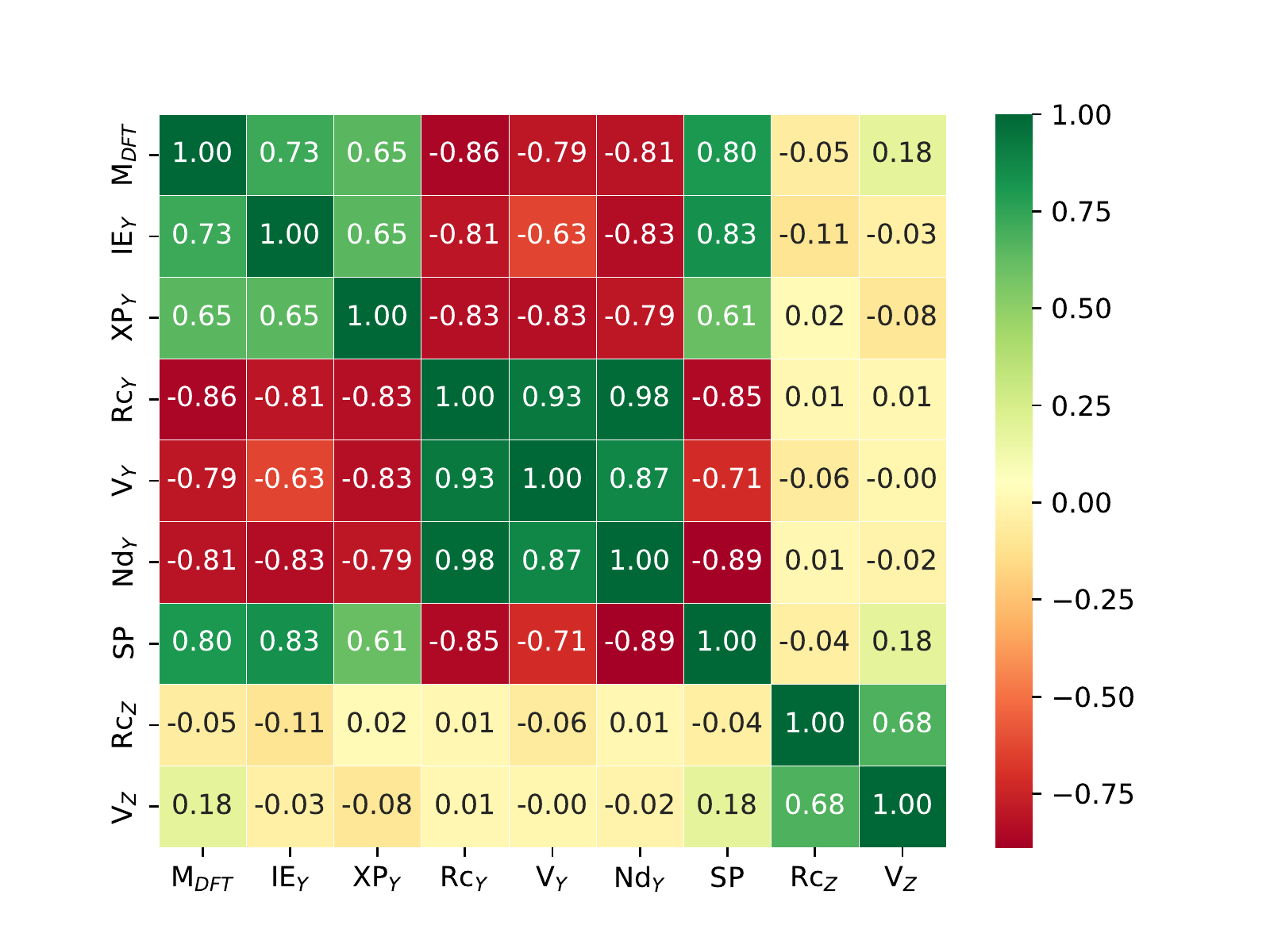}}
    \caption{Correlation between the DFT-obtained magnetic moment data with different features (a) for the undoped Heusler compounds (b)considering both doped and undoped compounds.}
    \label{Heatmap1}
\end{figure}

To build up a machine learning model to predict the magnetic moment we need to select the appropriate features. As for pure Co$_2$ based Heusler compounds the magnetic moment can be determined from the SP rule, $M_{S}=N_v-24$, we consider this itself as a feature. The other features we consider are: the first ionization energy (IE$_Y$) at the Y-site, Pauling electronegativity at Y-site (XP$_Y$), the covalent radius of the Y-site (Rc$_Y$), atomic volume of the Y-site (V$_Y$) and finally the number of unoccupied d-electrons at the Y-site (Nd$_Y$). The target feature being the DFT-obtained magnetic moments, $M_{DFT}$. To build up a machine learning model with these features, particularly when the Y-site is doped with transition metals we sought the help of Pearson correlation co-efficient which is given by,
\begin{equation}
  r (x,y) =
  \frac{ \sum_{i=1}^{n}(x_i-\bar{x})(y_i-\bar{y}) }{%
        \sqrt{\sum_{i=1}^{n}(x_i-\bar{x})^2}\sqrt{\sum_{i=1}^{n}(y_i-\bar{y})^2}}
\label{pearson}        
\end{equation}
which is a measure of the linear correlation between two variables $x$ and $y$, 
where $x,y \in \lbrace IE_Y,XP_Y,Rc_Y,V_Y,\\ Nd_Y,M_{SP},M_{DFT}\rbrace$. If we look at the Fig.\ref{Heatmap1} (a) where all Pearson correlation for only the undoped compounds are reported, we see $M_{DFT}$ is perfectly described by the $M_{SP}$, the value of $r$ being 1.  However, our goal is to identify a descriptor that can be uniformly used for samples that are doped, undoped, with, or without half-metallic features. For this purpose, we show the Pearson correlation for dataset containing both the doped and undoped samples are present in Fig.\ref{Heatmap1} (b). In this case, we see that the most important descriptors are $M_{SP}$, Rc$_Y$ and Nd$_Y$. But as Rc$_Y$ and Nd$_Y$ are highly correlated ($r\sim 0.98$) we have to consider only one among them. Considering the fact that the magnetic moment from a DFT calculation is given by: $\mu_B \times (\text{Number of up electrons}-\text{Number of down electrons})$ which has somewhat similarity to the number of unoccupied d-electrons, Nd$_Y$, we choose Nd$_Y$ as a feature to fit the magnetic moment data with machine learning models in addition to $M_{SP}$. The primary feature space $\Phi_0=\{M_{SP},Nd_Y\}$ therefore contains only two features. We consider up to four iterations to construct the feature spaces $\Phi_1$, $\Phi_2$, $\Phi_3$ and $\Phi_4$ and generate a total of 9006 features. In supplemental materials we report all the three descriptors (up to dimension three) that we obtain with the SISSO-approach. In the table\ref{Errors}, we show the errors predicted by these three models. It can be seen that all the metrics, RMSE, R$^2$ and MaxAE (maximum absolute error) are better for for all three SISSO-models in comparison to Slater-Pauling rule. As the data set is quite small, we check the validity of this model with leave-one-out cross validation (LOOCV). We find that both the Root Mean Square Error (RMSE) and Maximum Absolute Error (MaxAE) does not change in LOOCV.\\
In the following, we develop an effective model, which we refer to as a general rule, based on the SISSO descriptor of dimension three, and use it to predict additional Heusler compounds created by Y-site doping.
\subsection{Formulation of a general rule for predicting the magnetic moments of Co$_2$YZ \& Co$_2$Y$_{1-x}$M$_x$Z compounds:}
In the supplemental material, we show all the three descriptors we obtained. 
If we define the ratio $r=\frac{M_{SP}}{Nd_Y}$, we can simplify the 3D descriptor (supplemental materials, Table-I) by ignoring the second term which has very small coefficient ($\sim 10^{-5}$).
We also ignore the fact that the coefficients in front of the first and last term are slightly bigger than unity. We round them up to 1, as this allows to reformulate the our new rule in an compact fashion.
\begin{equation}
M'_{SP}=Nd_Y(1-e^{-r^3})+M_{SP}e^{-r}
\label{msp}
\end{equation}
Although this  \textit{new rule} gives slightly higher error with respect to the original 3D SISSO descriptor, it is more handy to use. In the table \ref{doped}, we show the magnetic moments of Co$_2$Y$_{1-x}$M$_x$Z compounds obtained from SP rule as well our descriptors. Finally both the approaches are compared with \textit{ground truth} values: which in this case are the ones obtained from the DFT calculations are made. In all cases, the results are shown for the case $x=0.25$. The alloys are still cubic  with space group $Pm3m (\text{no}~ 221)$. One would therefore expect similar kind of crystal field split of the d-levels. We define two metrics $\delta{SP}=M_{DFT}-M_{SP}$ and $\delta M'_{SP}=M_{DFT}-M'_{SP}$ using which we demonstrate the robustness of our proposed descriptor. Here $M_{DFT}$ is the magnetic moment obtained by DFT and $M'_{SP}$ is the magnetic moment obtained by our model.
It can be seen that for doping with the early transition metals such as Cr,V and Ti, the magnetic moments obtained from both $M_{SP}$ and $M'_{SP}$ are almost same and very close to the results obtained via DFT. But as soon as we introduce the late transition metals such as Ni, Cu and Zn, $M_{SP}$ fails miserably. The descriptor proposed by us, $M'_{SP}$ gives much better results in these cases. Our main result is that: The number of unoccupied electrons plays an important role. To account the magnetic moments properly particularly when the doping is done with late transition metal, the original Slater-Pauling rule has to be corrected. The correction should involve the number of unoccupied d-electrons in the dopant site.
It is worth to mention here that to account the magnetic moments of the half-metallic Heusler compounds with Y-site occupied by late TM atoms, Skaftouros \textit{et al.}~\cite{skaf} has proposed the so called generalized Slater-Pauling rules which states that for the case when Y site is occupied by a late TM atom, SP rule would read $M_{SP}=N_v-28$. This criterion, however, only applies when the late TM has entirely filled the Y site. In the current study, we can see that this rule does not work for  Co$_2$Zn$_{0.25}$Mn$_{0.75}$P or Co$_2$Cu$_{0.25}$Mn$_{0.75}$P. This can be understood by looking the table-\ref{doped}. 

\begin{figure*}[h!]
\includegraphics[scale=0.85]{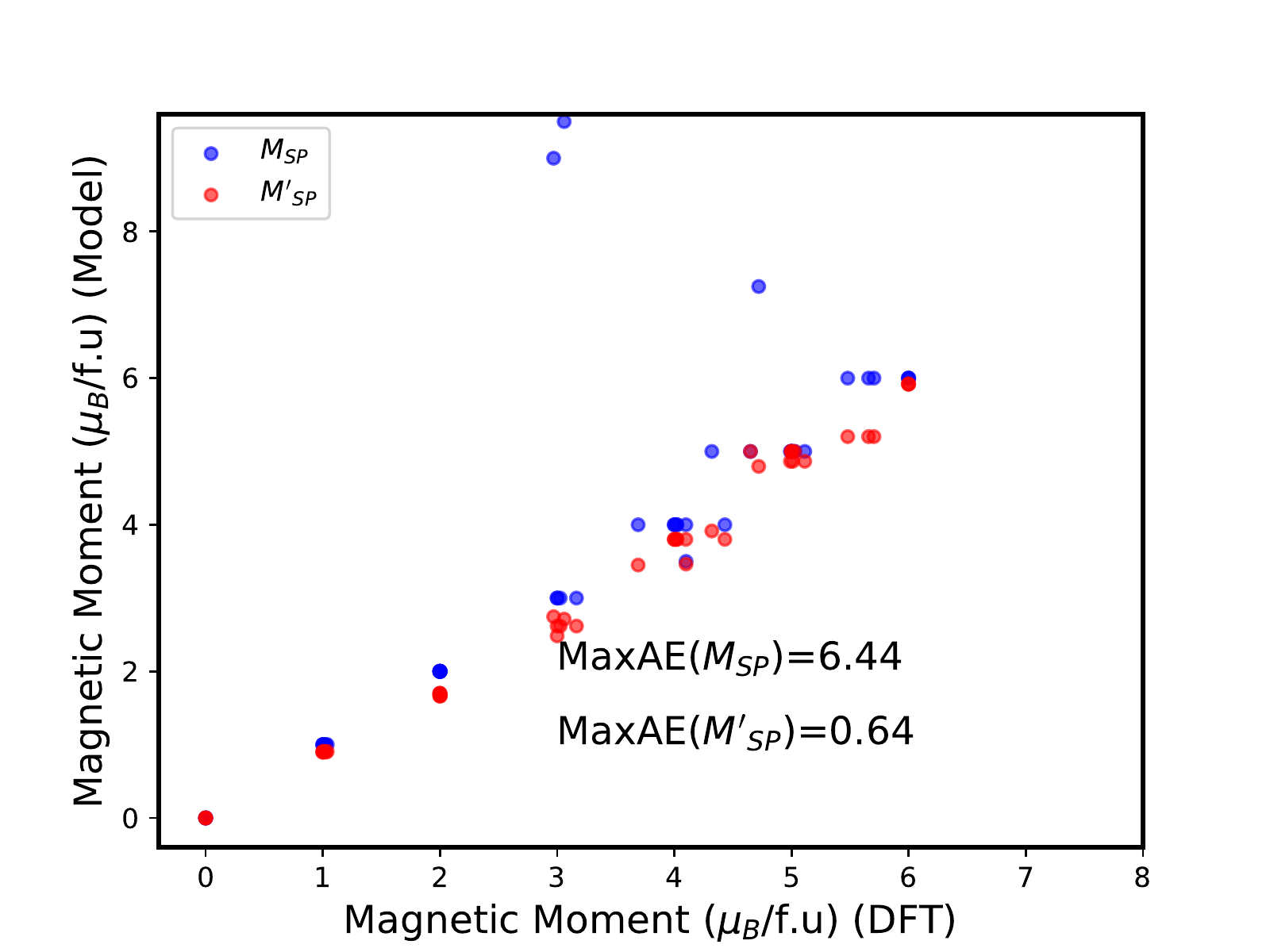}
\caption{Magnetic moments obtained different models such as Slater-Pauling ($M_{SP}$) and the general rule ($M'_{SP}$) are compared with the DFT-obtained magnetic moment data.}
\label{Comp}
\end{figure*}

\begin{figure*}[h!]
\includegraphics[width=0.75\columnwidth,height=120mm]{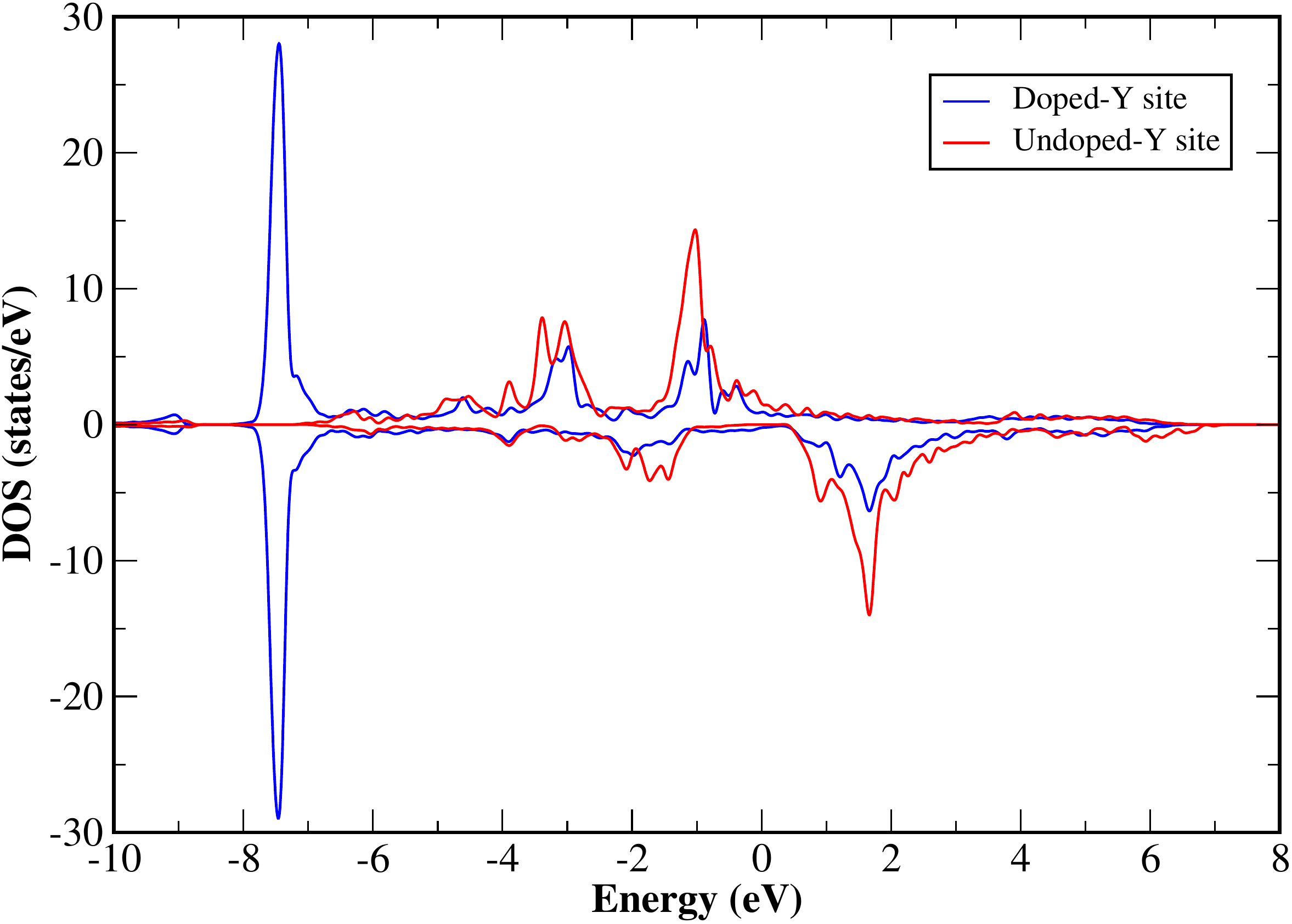}\caption{ Density of states of the Y site for Co$_2$MnSi and Co$_2$Zn$_{0.5}$Mn$_{0.5}$Si (in both cases, the number of atoms are same). The Fermi energy is at 0 eV.}
\label{TT1}
\end{figure*}

In the Fig.\ref{Comp}, we show the correlation plots between the DFT-obtained magnetic moment and the ones that obtained from the model. Alongside with the data shown in the table-\ref{doped}, it can be understood that the standard SP rule fails badly for the materials like Co$_2$Zn$_{0.5}$Mn$_{0.5}$Si and Co$_2$Cu$_{0.5}$Mn$_{0.5}$Sn, while the general one works pretty well. It can be also that even $M_{SP}=N_v-28$ does not suit well for these compounds. To understand how simply counting the number of valence electrons is not sufficient we compare  the density of states (DOS) of the Y-sites of both Co$_2$MnSi  and Co$_2$Zn$_{0.5}$Mn$_{0.5}$Si in the Fig.\ref{TT1}. As can be seen (also quite expected), the DOS exhibits characteristics that are a result of more than just the rigid shift caused by an increase in the number of valence electrons. Particularly if we look the DOS of the unoccupied part of the Y-site for the undoped case (total DOS of the 4 Mn atoms of the conventional cell) to the doped case (total DOS of the 4 Mn atoms and 4 Zn atoms), this can be seen clearly.
It is interesting to note here that, if we look at the total DOS (see the supplemental material), it can be seen that Co$_2$Zn$_{0.5}$Mn$_{0.5}$Si is almost half-metallic, but neither $M_{SP}=N_v-24$ nor $M_{SP}=N_v-28$ work for it. 
The worst performance of the new model is for Co$_2$Cu$_{0.25}$Mn$_{0.75}$Sn where the actual moment is 4.2$\mu_B$, while the predicted value is about 3.5 $\mu_B$, which is however close to the one obtains with SP rule. This corresponds to the maximum absolute error of 0.64 $\mu_B$ as shown in the Fig.\ref{Comp}.
\par
From the  Eq.\ref{msp}, it can be seen that for  when $Nd_Y$ is very large, $r$ is very small (close to zero) and $M'_{SP} \longrightarrow M_{SP}$ (See also the supplemental materials, Figure-1). This is the case when Y-site does not contain any late TM. The conventional SP rule works well even when doping is done with early transition metals such as Ti or Cr as can be seen from the table-\ref{doped}. 

\subsection{Stability check for the predicted materials}
Finally, we check the doped materials which are proposed in the table-\ref{doped}, are they stable or not. For that we again use the compressed sensing method described above. This time, we consider the whole data set, i,e both L$_{21}$  and $D022$. As we have non-cubic materials as well, we use more number of primary features to train the model. We choose six primary features: atomic number(Z), ionization energy(IE), the Pauling electro-negativity($\chi$), covalent radius ($r$), the number of valence electrons ($N$) and the atomic volume. In this particular case we use both L$_{21}$ and D$022$ database of formation energies. 

\begin{figure*}[h!]
\includegraphics[width=0.75\columnwidth,height=120mm]{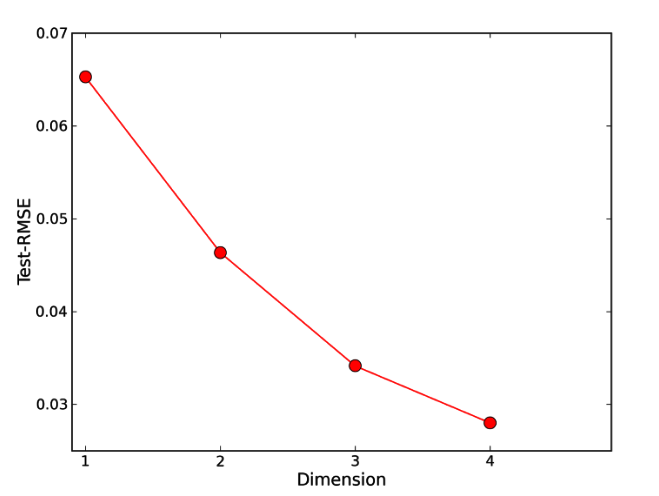}
\caption{Test-RMSE of the six-fold cross validation with the dimension of the descriptor for the formation energy.}
\label{RMSE}
\end{figure*}

\begin{figure*}[h!]
\includegraphics[width=0.75\columnwidth,height=120mm]{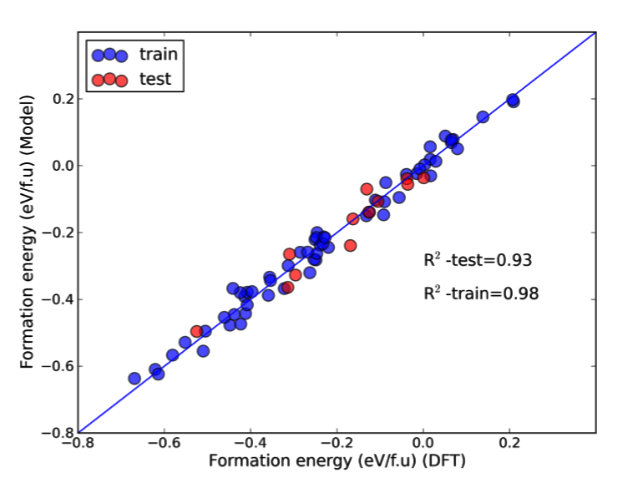}
\caption{ Performance of the SISSO-learned 4D descriptor for the formation energy.}
\label{TT}
\end{figure*}

We employ a somewhat higher dimensional descriptor to accurately predict the formation energy. This time, we constructed a four dimensional descriptor. Because a higher dimensional descriptor is employed, there is a risk of overfitting, thus we undertake cross validation (CV). We split the data into test and training set and perform a six-fold cross validation. $80\%$ of the data were used for training and $20\%$ for the testing.
In the Fig.\ref{RMSE}, we show the root mean square error (RMSE) for test-CV. The error decreases with the dimension and for the four dimensional descriptor the CV-error is small enough (0.03 eV/f.u). In the Fig.\ref{TT}, we show the performance of the above descriptor in terms of describing the DFT data set. The descriptor works well with the DFT values. We can see that R$^2$ values for the training and testing sets are very close by showing the effectiveness of the model. 

In table-\ref{stability}, we compare the stability of a few doped Heusler alloys as predicted by DFT to what we expected using the four-dimesnional descriptor that was obtained as explained above. The DFT-formation energies are calculated via~\cite{form1,quart},
\begin{equation}
\Delta E_f=E_{Co_2Y_{1-x}M_xZ}- \frac{1}{4}(2\mu_{Co}+(1-x)\mu_{Y}+x\mu_M+\mu_Z)
\label{form}
\end{equation}
In the above equation, the first term is the total energy of the Heusler compound Co$_2$Y$_{1-x}$M$_x$Z while the second term consists of the sum of the chemical potentials of its various constituents. 

In the case of the doped compounds, it is important to provide the results for just those with appreciable magnetic moments, thus we chose the compounds based on that criterion. We choose Co$_2$MnP as the parent compound and consider the cases when fractional substitution by other transition metals that lead to significant moment. We see that the compounds predicted in the table-\ref{doped} are quite stable from the point view of the formation energy calculated by DFT-based approach as well as from the compressed sensing method. 
\section{Conclusions}
In conclusion, we suggest a general rule for forecasting the magnetic moments of Co$_2$-based full-Heusler alloys. Using online magnetic moment data for Co$_2$-based Heusler compounds and a compressed sensing technique, we arrived at a magnetic moment descriptor that depends not only on the number of valence electrons in the unit cell, but also on the number of unoccupied d-electrons of the doping site. We show that this formula works well for both regular L2$_1$ Heusler compounds and doped Heuslers with slightly lower symmetry ($Pm\bar{3}m$). Most importantly, unlike Slater-Pauling rule which is accurate only at the half-metallic limit, this new rule works for any  Co$_2$-based Heusler compound. We show that the  performance of this new rule is quite impressive in the case when the Heusler compound is doped with the late transition metals.
.


\section{Acknowledgement}
We thank Prof. Iosif Galanakis, University of Patras, for the valuable comments. This work was supported by NRF grant funded by MSIP, Korea(No. 2009-0082471 and No. 2014R1A2A2A04003865),the Convergence Agenda Program (CAP) of the Korea Research Council of Fundamental Science and Technology (KRCF) and GKP (Global Knowledge Platform) project of the Ministry of Science, ICT and Future Planning.
\newpage
\clearpage
\bibliography{Heus}
\newpage

\begin{table}[ht]
	\centering
\begin{tabular}{c|c|c|c}
     Method & RMSE & R$^2$ & MaxAE \\
     \hline
     Slater-Pauling & 1.35 & 0.68 & 6.4 \\
     \hline
     SISSO-model-1D & 0.5 & 0.92 & 1.9\\
     \hline
     SISSO-model-2D & 0.35 & 0.96 & 0.76\\
     \hline
     SISSO-model-3D & 0.19& 0.99 & 0.58\\
     \hline
     \hline
     \textbf{General rule} & 0.35 & 0.96 & 0.64\\
     \hline
\end{tabular}
\caption{Comparison of errors for the magnetic moments obtained by different models: Slater-Pauling, SISSO-models of different dimesnions and finally a 2D model model which is referred as general rule.}
\label{Errors}
\end{table}
\newpage

\begin{table}[ht]
	\centering
	\small
\setlength{\tabcolsep}{4pt}
		\begin{tabular}{c c c c c c c}
			\toprule
			\textbf{Compound} &  \textbf{M$_{SP}$} & \textbf{$M'_{SP}$} & \textbf{M$_{DFT}$} & $\delta M_{SP}$ & $\delta M'_{SP}$\\
			\hline
			\midrule
			\hline
	        Co$_2$Ti$_{0.25}$Mn$_{0.75}$ P & 5.25 & 5.20 & 5.11 & -0.14 &-0.09  \\
	        \hline
	        Co$_2$Cr$_{0.25}$Mn$_{0.75}$ P & 5.75 & 5.73 & 5.77& -0.02 & 0.04 \\
	        \hline
	        Co$_2$V$_{0.25}$Mn$_{0.75}$ P & 5.50 & 5.50 & 5.50& 0.00 & 0.00\\
	        \hline
			Co$_2$Fe$_{0.25}$Mn$_{0.75}$ P & 6.25 & 5.93 & 6.15 & -0.10& 0.20 \\
			\hline
	        Co$_2$Ni$_{0.25}$Mn$_{0.75}$ P & 6.75 & 5.55 & 4.86 & -1.89 & -0.69\\
	        \hline
	        Co$_2$Cu$_{0.25}$Mn$_{0.75}$ P & 7.00 & 5.20 & 4.50 & -2.50 & -0.69 \\
	        \hline
	        Co$_2$Zn$_{0.25}$Mn$_{0.75}$ P & 7.25 & 4.8 & 4.75 & -2.50 & -0.46 \\
	        \hline
	        \hline
	        Co$_2$Ti$_{0.25}$Mn$_{0.75}$ As & 5.25 & 5.21 & 5.23 & -0.05 & -0.04 \\
	        \hline
	        Co$_2$Zn$_{0.25}$Mn$_{0.75}$ As & 7.25 & 4.73 & 4.73 & -2.52 & 0.00\\
	        \hline
	        \hline
	        Co$_2$Ti$_{0.25}$Mn$_{0.75}$ Sb & 5.25 & 5.20 & 5.21 & -0.04 &-0.01 \\
	        \hline
	        Co$_2$Zn$_{0.25}$Mn$_{0.75}$ Sb & 7.25 & 4.72 & 4.79 & -2.46 & -0.07\\
	        \hline
	        \hline
	        \bf{Co$_2$Zn$_{0.5}$Mn$_{0.5}$Si} & \textbf{7.5} & \textbf{2.71} & \textbf{3.06} & \textbf{-4.44} & \textbf{-0.35}\\
	        \hline
	        \bf{Co$_2$Cu$_{0.5}$Mn$_{0.5}$Sn} & \textbf{7.0} & \textbf{2.75} & \textbf{2.97} & \textbf{-4.03} & \textbf{-0.22}\\
			\bottomrule
		\end{tabular}
	\caption{Comparison between the magnetic moments obtained from the different approaches for Co$ 2$Y$_{1-x}$M$_x$Z Heusler alloys. Here $M_{SP}$ refers to the standard Slater-Pauling rule ($M_{SP}=N_v-24$).}
	\label{doped}
	\end{table}
	
\begin{table}[ht]
	\centering
	\small
\setlength{\tabcolsep}{4pt}
		\begin{tabular}{c c c c c}
			\toprule
			\textbf{Compound} & \textbf{M$_{DFT}$ ($\mu_B$)} &\textbf{Formation energy (DFT)(eV/atom)} & \textbf{Formation energy (SISSO)(eV/atom)} \\
			\hline
			\midrule
			\hline
			Co$_2$Fe$_{0.25}$Mn$_{0.75}$ P & 6.15 & -0.23&-0.33 \\
			\hline
	        Co$_2$Ti$_{0.25}$Mn$_{0.75}$ P &5.11 &-0.31& -0.38\\
	        \hline
	        Co$_2$Cr$_{0.25}$Mn$_{0.75}$ P& 5.77 &-0.23 & -0.32 \\
	        \hline
	        Co$_2$V$_{0.25}$Mn$_{0.75}$ P & 5.50 & -0.34 & -0.35 \\
			\bottomrule
		\end{tabular}
	\caption{Formability of some selected Co$ 2$Y$_{1-x}$M$_x$Z Heusler alloys. A comparison with DFT is also illustrated.}
	\label{stability}
	\end{table}
\clearpage
\newpage
\clearpage
\newpage
\newpage
\clearpage
\newpage
\clearpage
\newpage
\newpage
\clearpage
\newpage
\clearpage
\end{document}


\begin{table}[ht]
	\centering
\begin{tabular}{c|c|c|c}
\hline
     Dimesnion & Descriptor & SIS& \\
     \hline
     1 & $(M_{SP}/exp((M_{SP}/exp(Nd_Y))))$ & 100\\
     \hline
     2& $1.2*(M_{SP}/exp((M_{SP}/(Nd_Y)^2)))+0.007*((M_{SP}/((Nd_Y-M_{SP})-M_{SP})))^3$ & 200\\
     \hline
     3 & $1.03*(Nd_Y+(M_{SP}/exp((M_{SP}/Nd_Y))))+0.00003*exp((Nd_Y/(M_{SP}+(M_{SP}-Nd_Y))))$&300\\
     & $-1.01*(Nd_Y/exp(((M_{SP}/Nd_Y))^3))$ & \\
     \hline
\end{tabular}
\caption{Descriptors identified by SISSO with error analysis for the case of magnetic moment}
\label{Errors}
\end{table}
\newpage
\begin{figure*}[h!]
\includegraphics[width=0.75\columnwidth,height=120mm]{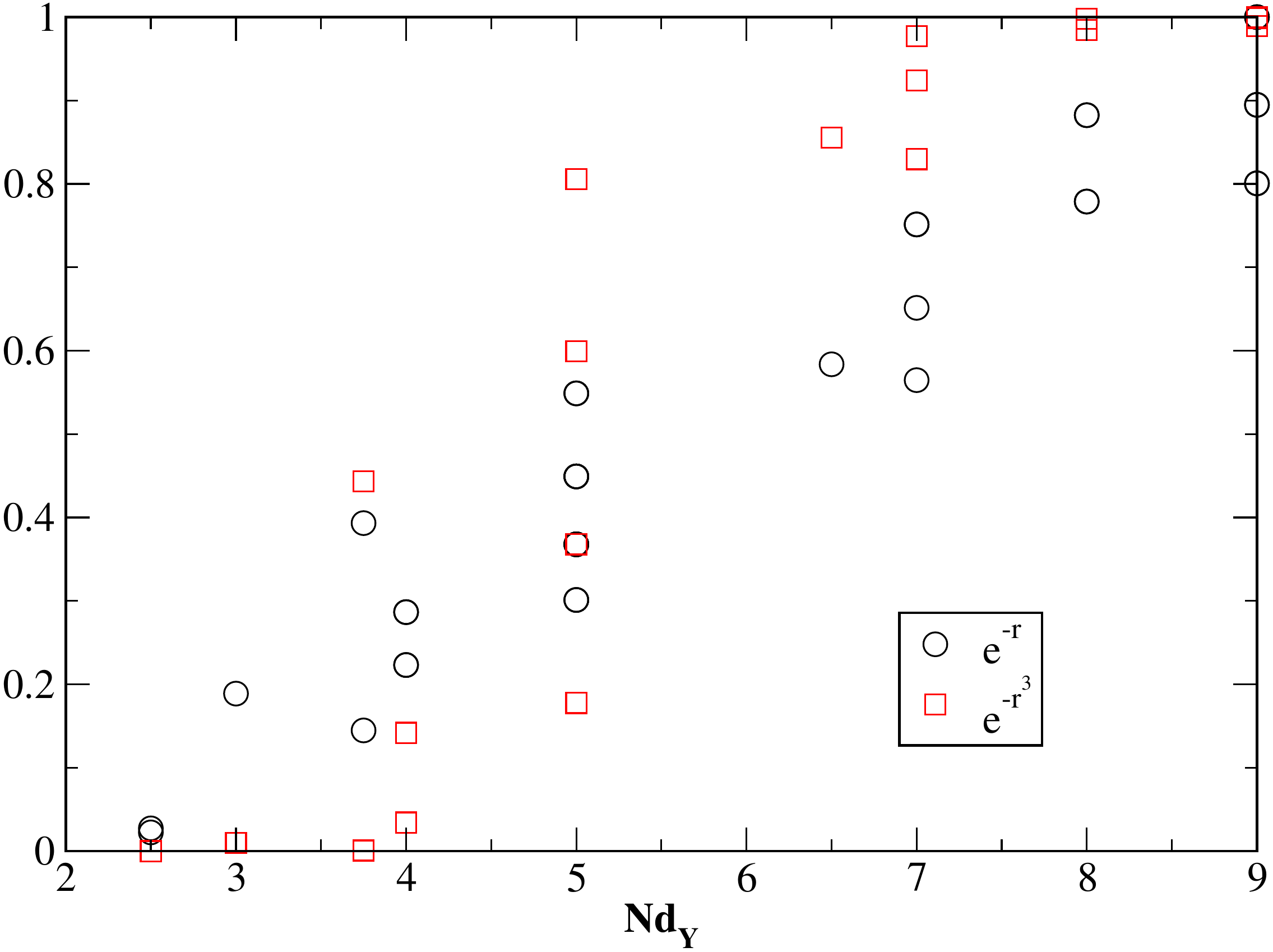}
\caption{ Plot of the exponential factors: $e^{-r}$ and $e^{{-r}^3}$ used in the Eqn.3 of the manuscript. $r=\frac{M_{SP}}{Nd_Y}$. It can be seen that the exponential factors are close to 1 for small values of $r$ (large values of $Nd_Y$) giving to $M'_{SP}\sim M_{SP}$ for early TM doped compounds.}
\label{TT}
\end{figure*}  
\newpage
\begin{figure*}[h!]
\includegraphics[width=0.75\columnwidth,height=120mm]{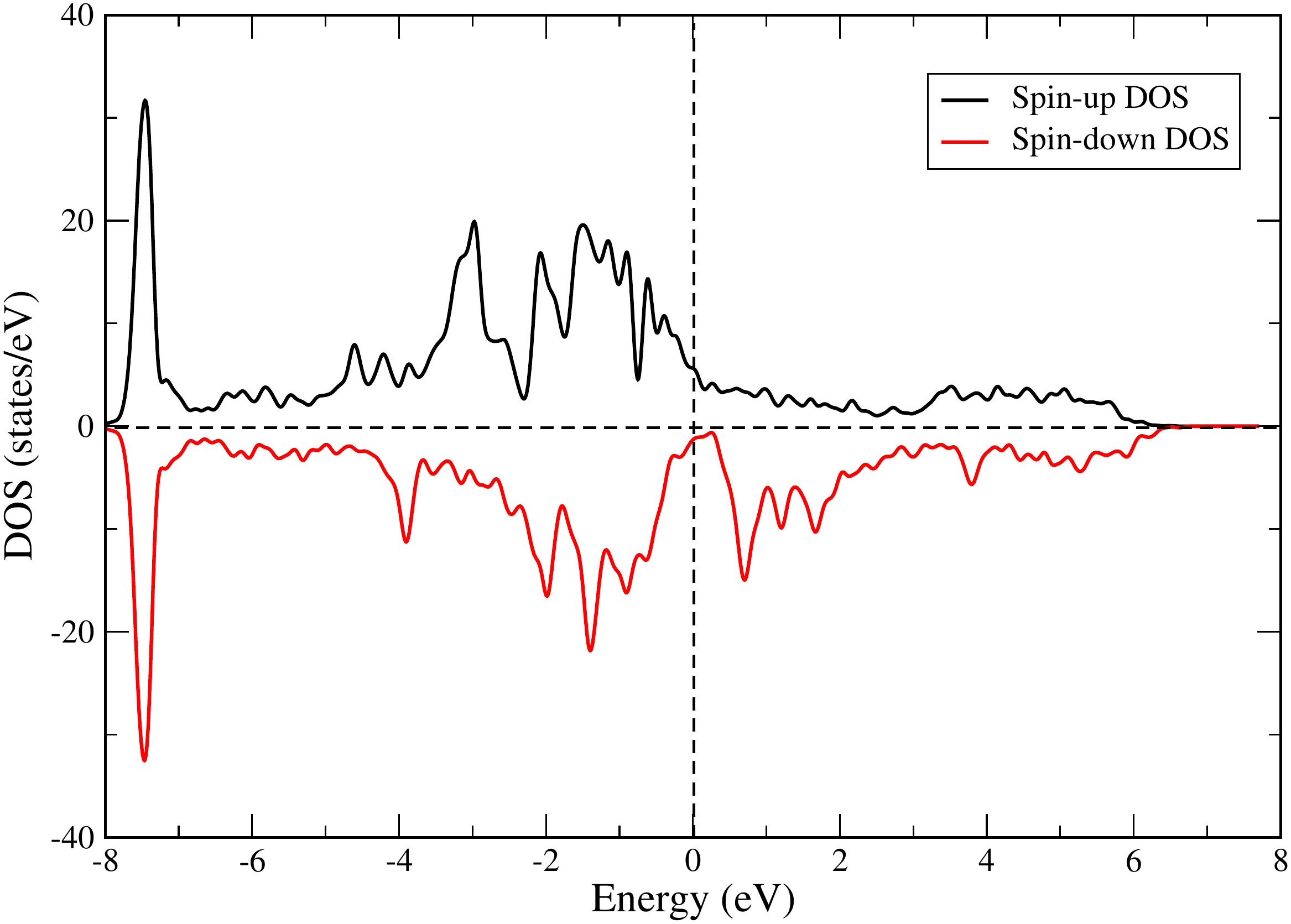}
\caption{ Total density of states of Co$_2$Zn$_{0.5}$Mn$_{0.5}$Si. It can be seen that the material is almost half-metallic, yet neither version of SP rule work for it.}
\label{TT1}
\end{figure*}